\documentclass[journal,10pt]{IEEEtran}
\pdfoutput=1
\usepackage{times,amsmath,epsfig}
 \usepackage{algpseudocode}
\usepackage{algorithm}
\usepackage{algorithmicx}
\usepackage{enumitem}
\usepackage{chemarrow}
\usepackage{graphicx}
\usepackage{stfloats}
 \usepackage{cite}
 \usepackage{color}
 \usepackage{setspace}
 \usepackage{psfrag}
 \usepackage{subfigure}
 \usepackage{amssymb}
 \usepackage{epsfig}
 \usepackage{pifont}
 \usepackage{amsmath}
 \allowdisplaybreaks[4]
 \usepackage{flushend}
 \usepackage{array}
 \usepackage{multicol}
 \usepackage{amsfonts}
 \usepackage{lipsum} 
 \usepackage{multirow}
 \usepackage{verbatim}
 \usepackage{rotating}
\usepackage[acronym]{glossaries}
\usepackage{lscape}
 \usepackage{makecell}
 \usepackage[color]{changebar}
 \cbcolor{red}
 \usepackage[switch]{lineno}
 \usepackage{mathtools}

\newcommand{\argmin}{\arg\,\min}

\begin{document}
\title{Low Complexity Detection of Spatial Modulation Aided OTFS in Doubly-Selective Channels}

\author{{Zeping Sui}, {\em Student Member,~IEEE}, {Hongming~Zhang}, {\em Member,~IEEE}, {Yu Xin}, {Tong Bao}, {Lie-Liang Yang}, {\em Fellow,~IEEE}, and {Lajos Hanzo}, {\em Life Fellow,~IEEE}
\thanks{This work was supported by the China Scholarship Council under Grant 202004910653 and the National Natural Science Foundation of China under Grant 62001056. (\emph{Corresponding author: Hongming Zhang})}
\thanks{Zeping Sui is with the Institute of Acoustics, Chinese Academy of Sciences, Beijing 100190, China; and also with the University of Chinese Academy of Sciences, Beijing 100049, China (e-mail: suizeping@mail.ioa.ac.cn).}
\thanks{Hongming Zhang is with the School of Information and Communication Engineering, Beijing University of Posts and Telecommunications, Beijing 100876, China (e-mail: zhanghm5685@163.com).}
\thanks{Yu Xin and Tong Bao are with the State Key Laboratory of Mobile Network and Mobile Multimedia Technology, ZTE Corporation, Shenzhen 518055, China (e-mail: xin.yu@zte.com.cn; bao.tong@zte.com.cn).}
\thanks{Lie-Liang Yang and Lajos Hanzo are with the Department of Electronics and Computer Science, University of Southampton, Southampton SO17 1BJ, U.K. (e-mail: lly@ecs.soton.ac.uk; lh@ecs.soton.ac.uk).}%
\thanks{L. Hanzo would like to acknowledge the financial support of the Engineering and Physical Sciences Research Council projects EP/W016605/1 and EP/X01228X/1 as well as of the European Research Council's Advanced Fellow Grant QuantCom (Grant No. 789028)}
}
\maketitle

\begin{abstract}
A spatial modulation-aided orthogonal time frequency space (SM-OTFS) scheme is proposed for high-Doppler scenarios, which relies on a low-complexity distance-based detection algorithm. We first derive the delay-Doppler (DD) domain input-output relationship of our SM-OTFS system by exploiting an SM mapper, followed by characterizing the doubly-selective channels considered. Then we propose a distance-based ordering subspace check detector (DOSCD) exploiting the \emph{a priori} information of the transmit symbol vector. Moreover, we derive the discrete-input continuous-output memoryless channel (DCMC) capacity of the system. Finally, our simulation results demonstrate that the proposed SM-OTFS system outperforms the conventional single-input-multiple-output (SIMO)-OTFS system, and that the DOSCD conceived is capable of striking an attractive bit error ratio (BER) vs. complexity trade-off.
\end{abstract}
\IEEEpeerreviewmaketitle
\vspace{-0.9em}
\section{Introduction}\label{Section1}
Spatial modulation (SM) has evolved into a compelling multiple-input multiple-output (MIMO) technique~\cite{6678765,4382913,9729571}, where the information is jointly conveyed by the classic amplitude-phase modulated (APM) symbols and the indices of the activated transmit antennas (TAs), yielding a high energy-efficiency (EE). A suite of low-complexity SM detectors were proposed in~\cite{6679367,6376045,6784532,9468877}, such as lattice-based (LB)~\cite{6679367}, distance-based (DB)~\cite{6376045} and ordering-based (OB) detectors~\cite{6376045,6784532,9468877}. However, these SM detectors have not been conceived for multicarrier (MC) communications.
As a parallel development, orthogonal time frequency space (OTFS) modulation has also matured into a promising delay-Doppler (DD)-domain modulation candidate for next-generation wireless networks~\cite{7925924,9508932,9508141,8686339,yuan2021iterative}. The SM-aided OTFS (SM-OTFS) concept has been proposed in~\cite{9521176,9718240}. However, both the inter-carrier interference (ICI) and inter-symbol interference (ISI) imposed by doubly-selective channels were ignored in~\cite{9521176}, while the time-domain detector of~\cite{9718240} disregarded the specific characteristics of the DD-domain doubly-selective (DDS) channels. More recently, the message passing detector (MPD) which is originally conceived for OTFS in \cite{8424569} has been extended to OTFS combined both with index modulation (OTFS-IM) and SM-OTFS systems \cite{10039709}.

Against this backdrop, we conceive a low-complexity detection aided SM-OTFS system for transmission over high-mobility channels. The contributions of our paper are boldly and explicitly contrasted to the literature in Table~\ref{table1}, which are further detailed below.
\begin{itemize}
	\item The DD-domain input-output relationship of the SM-OTFS system is derived, so as to take full advantage of the DDS channels and the SM properties.
	\item A novel low-complexity near-maximum likelihood (ML) distance-based ordering subspace check detector (DOSCD) is conceived, in which the reliabilities of different TA activation patterns (TAPs) are quantified, and the \emph{a priori} information of the transmit symbol vector is utilized for separately detecting the APM symbols and TAPs.
	\item We derive the discrete-input continuous-output memoryless channel (DCMC) capacity for characterizing the system performance. Furthermore, we benchmark the proposed detector against the ML detector (MLD) by simulations, illustrating that the DOSCD is capable of striking a compelling bit error ratio (BER) vs. complexity trade-off.
	\end{itemize}
\begin{table}[t]
\footnotesize
\centering
\caption{Contrasting Our Contributions to the Existing Literature}
\label{table1}
\begin{tabular}{l|c|c|c|c|c}
\hline
Contributions & \textbf{Our work} & \cite{6376045} & \cite{6784532,9468877} & \cite{9521176,9718240} & \cite{10039709}\\
\hline
\hline
SM-OTFS & \checkmark &  &  & \checkmark & \checkmark \\
\hline
DDS channels & \checkmark &  &  &  & \checkmark \\
\hline
MC communications & \checkmark &  &  & \checkmark & \checkmark \\
\hline
DB detection & \checkmark & \checkmark &  &  &  \\
\hline
OB detection & \checkmark & \checkmark & \checkmark &  &  \\
\hline
Capacity analysis & \checkmark &  &  &  &  \\ 
\hline
\end{tabular}
\end{table}

The rest of the paper is organized as follows. Section \ref{Section2} derives the system model of SM-OTFS, while we detail the proposed DOSCD and the DCMC capacity in Section \ref{Section3}. Our simulation results are provided in Section \ref{Section4}, followed by our conclusions in Section \ref{Section5}.
\section{SM-OTFS System Model}\label{Section2}
\begin{figure}[t]
\centering
\includegraphics[width=\linewidth]{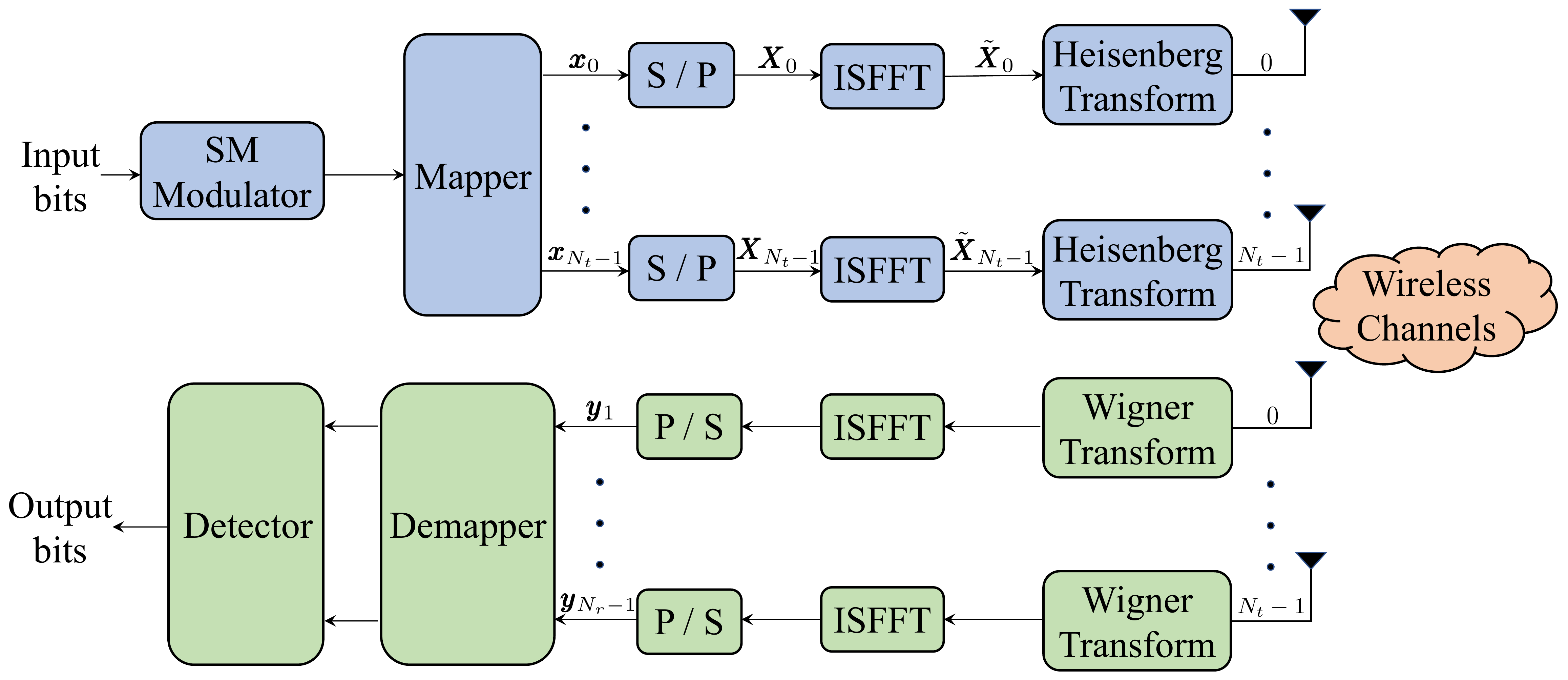}
\caption{The transceiver diagram of the SM-OTFS system.}
\label{Figure1}
\vspace{-2em}
\end{figure}

Consider a limited-dimensional MIMO-OTFS system having $N_t$ TAs and $N_r$ receive antennas (RAs), where the high-mobility TAs and RAs are randomly located in a wide area \cite{9590508,9906092,bjornson2015massive}. Let $\Delta f$ and $T$ denote the subcarrier frequency spacing and symbol duration, respectively. Hence, the bandwidth and frame duration of OTFS signals are given by $B=M\Delta f$ and $T_f=NT$, respectively, where $M$ is the number of subcarriers and $N$ denotes the number of time-slots (TSs) per frame. As shown in Fig. \ref{Figure1}, for information transmission, we first divide the length-$L_b$ input bits $\pmb{b}\in\mathbb{B}^{L_b}$ into $M_d=NM$ groups, yielding $\pmb{b}=[\pmb{b}_1,\ldots,\pmb{b}_{M_d}]$, where $\mathbb{B}$ represents the bit set $\{0,1\}$. Hence, there are $L=L_b/M_d=L_1+L_2$ bits in each group. Then, the $m_d$th bit sequence $\pmb{b}_{m_d}$ is split into two subsequences denoted as $\pmb{b}_{m_d,1}$ and $\pmb{b}_{m_d,2}$ for $m_d=1,\ldots,M_d$. By mapping the bit sequence $\pmb{b}_{m_d,1}\in\mathbb{B}^{L_1}$ into a TAP, one of the $N_t$ TAs is activated to convey the APM symbols, where we have $L_1=\log_2 N_t$. Moreover, based on the $Q$-ary normalized quadrature amplitude modulation (QAM)/phase-shift keying (PSK) constellation $\mathcal{A}=\{a_1,\ldots,a_Q\}$, the remaining $L_2=\log_2 Q$ bits in $\pmb{b}_{m_d,2}\in\mathbb{B}^{L_2}$ are mapped into a classic APM symbol. Overall, the SM-OTFS system can be specified as $(N_t,N_r,Q)$, giving the total number of bits per frame by $L_b=M_dL=M_d\log_2 (N_tQ)$, and the rate conveyed by $R=\log_2 (N_tQ)$ bits/s/Hz. Consequently, the overall OTFS frame $\pmb{S}\in\mathbb{C}^{N_t\times M_d}$ can be formulated as
\begin{align}\label{Eq1}
	\pmb{S}=\begin{bmatrix}
		S(0,0) & \ldots & S(0,M_d-1)\\
		\vdots & \ddots & \vdots \\
		S(N_t-1,0) & \ldots & S(N_t-1,M_d-1)
	\end{bmatrix},
\end{align}
where each column has only a single non-zero element corresponding to the activated TA. 

Let $\pmb{S}_{n_t,:}$ denote the $n_t$th row of $\pmb{S}$, which is transmitted by the $n_t$th TA. Furthermore, let $\pmb{x}_{n_t}=\pmb{S}_{n_t,:}^T$. Then, the $n_t$th DD-domain symbol matrix can be expressed as $\pmb{X}_{n_t}=\text{vec}^{-1}(\pmb{x}_{n_t})\in\mathbb{C}^{N\times M}$, where $\text{vec}^{-1}(\cdot)$ is the inverse vectorization operator. Based on the inverse symplectic finite Fourier transform (ISFFT) \cite{7925924}, the elements of the time-frequency (TF)-domain symbol matrix can be obtained as
\begin{align}\label{Eq2}
	\tilde{{X}}_{n_t}(n,m)=\frac{1}{\sqrt{M_d}}\sum_{k=0}^{N-1}\sum_{l=0}^{M-1}X_{n_t}(k,l)e^{j2\pi\left(\frac{nk}{N}-\frac{ml}{M}\right)},
\end{align}
for $n=0,\ldots,N-1$ and $m=0,\ldots,M-1$. Consequently, upon exploiting the Heisenberg transform, the time-domain signal to be transmitted by the $n_t$-th TA can be expressed as
\begin{align}\label{Eq3}
	\tilde{s}_{n_t}(t)=\sum_{n=0}^{N-1}\sum_{m=0}^{M-1}\tilde{{X}}_{n_t}(n,m)g_{\text{tx}}(t')e^{j2\pi m\Delta ft'},	
	\end{align}
where $t'=(t-nT)$ and $g_{\text{tx}}(t)$ represents the transmit pulse for $n_t=0,\ldots,N_t-1$.

Let us consider a $P$-path time-varying multi-path Rayleigh fading channel, whose channel impulse response (CIR) between the $n_t$th TA and $n_r$th RA is expressed as \cite{9590508,9906092}
\begin{align}\label{Eq4}
	h_{n_r,n_t}(\tau,\nu)=\sum_{i=1}^{P}h_{i,n_r,n_t}\delta(\tau-\tau_i)\delta(\nu-\nu_i),
\end{align}
where $h_{i,n_r,n_t}$, $\tau_i$ and $\nu_i$ are the complex-valued path gain, delay- and Doppler-shifts associated with the $i$th path, and $\delta(\cdot)$ denotes the Dirac-delta function. According to \cite{7925924}, we have $\tau_i=\frac{l_i}{M\Delta f}$ and $\nu_i=\frac{k_i}{NT}$, where $l_i$ and $k_i$ denote the normalized delay and Doppler indices of the $i$th path. We note that if the RAs are hosted by a fixed-location base station (BS) in a uniform linear array (ULA) form, the channel model of \eqref{Eq4} may be refined as the delay, Doppler and angular (DDA)-domain channel model \cite{9891774,10042436}. Consequently, by considering the array steering vectors at both RA and TA sides, our proposed SM-OTFS scheme can be directly extended to DDA-domain channel-based MIMO-OTFS systems. Moreover, $h_{i,n_r,n_t}$ are the independent and identically distributed (i.i.d.) random variables that have a mean of zero and a variance of $1/P$, i.e., we have $h_{i,n_r,n_t}\sim\mathcal{CN}(0,1/P)$.

At the receiver side, the time-domain signal received by the $n_r$th RA from the $n_t$th TA can be formulated as $r_{n_r,n_t}(t)=\int\int h_{n_r,n_t}(\tau,\nu)\tilde{s}_{n_t}(t-\tau)e^{j2\pi\nu(t-\nu)}d\tau d\nu+n_{n_r,n_t}(t)$, where $n_{n_r,n_t}(t)$ is the complex additive white Gaussian noise (AWGN) obeying $\mathcal{CN}(0,\sigma^2)$. By leveraging the receive pulse $g_{\text{rx}}(t)$ and the Wigner transform, the elements of the TF-domain symbol matrix of the $n_r$th RA received from the $n_t$th TA can be formulated as
\begin{align}\label{Eq5}	
\tilde{Y}_{n_r,n_t}(n,m)=\int r_{n_r,n_t}(t)g_{\text{rx}}(t')e^{j2\pi m\Delta ft'}dt.
\end{align}
The elements of the corresponding DD-domain received symbol matrix can be expressed as
\begin{align}\label{Eq6}
	Y_{n_r,n_t}(n',m')=\sum_{n=1}^{N}\sum_{m=1}^{M}\frac{\tilde{Y}_{n_r,n_t}(n,m)}{\sqrt{M_d}}e^{-j2\pi\left(\frac{nk}{N}-\frac{ml}{M}\right)},
\end{align}
for $n'=1,\ldots,N$ and $m'=1,\ldots,M$. By stacking the columns of $\pmb{Y}_{n_r,n_t}$ to formulate a column vector, the corresponding DD-domain symbol vector can be obtained as $\pmb{y}_{n_r,n_t}=\text{vec}(\pmb{Y}_{n_r,n_t})$. We assume that bi-orthogonal transmit and receive pulses are used. Then, the $n_r$th DD-domain received symbol can be obtained by exploiting the vector-form input-output relationship of
\begin{align}\label{Eq7}
	\pmb{y}_{n_r}=\sum_{n_t=0}^{N_t-1}\pmb{y}_{n_r,n_t}=\sum_{n_t=0}^{N_t-1}\pmb{H}_{n_r,n_t}\pmb{x}_{n_t}+\pmb{n}_{n_r},
\end{align}
where $\pmb{n}_{n_r}$ is the DD-domain AWGN vector, and the DD-domain channel matrix $\pmb{H}_{n_r,n_t}$ can be expressed as \cite{yuan2021iterative}
\begin{align}\label{Eq8}
	\pmb{H}_{n_r,n_t}=\sum_{i=1}^{P}\pmb{I}_M(l_i)\otimes\left[\pmb{I}_N(k_i)h^{(u)}_{i,n_r,n_t}e^{-j2\pi\frac{l_ik_i}{M_d}}\right].\end{align}
Let $\pmb{x}=[\pmb{x}_0^T,\ldots,\pmb{x}_{N_t-1}^T]^T\in\mathbb{C}^{M_dN_t\times1}$ and $\pmb{n}=[\pmb{n}_0^T,\ldots,\pmb{n}_{N_r-1}^T]^T\in\mathbb{C}^{M_dN_r\times1}$ represent the transmit DD-domain stacked vector and the received stacked noise vector, respectively. Moreover, the DD-domain MIMO channel matrix $\pmb{H}\in\mathbb{C}^{M_dN_r\times M_dN_t}$ can be expressed as
\begin{align}\label{Eq8-1}
	\pmb{H}=\begin{bmatrix}
		\pmb{H}_{0,0} & \pmb{H}_{0,1} & \cdots & \pmb{H}_{0,N_t-1}\\
		\pmb{H}_{1,0} & \pmb{H}_{1,1} & \cdots & \pmb{H}_{1,N_t-1}\\
		\vdots & \vdots & \ddots & \vdots\\
		\pmb{H}_{N_r-1,0} & \pmb{H}_{N_r-1,1} & \cdots & \pmb{H}_{N_r-1,N_t-1}			
		\end{bmatrix}.
\end{align}
Therefore, the DD-domain end-to-end input-output relationship can be formulated as
\begin{align}\label{Eq9}
	\pmb{y}=\pmb{H}\pmb{x}+\pmb{n},
\end{align}
where $\pmb{y}=[\pmb{y}_0^T,\ldots,\pmb{y}_{N_r-1}^T]^T\in\mathbb{C}^{M_dN_r\times1}$ denotes the received stacked vector. 

As shown in Fig. \ref{Figure1}, to exploit the properties of SM, we introduce a $(M_dN_t\times M_dN_t)$-dimensional SM mapper matrix $\pmb{\Upsilon}$, which is referred to as the \emph{perfectly shuffled} row-column in \cite{van2000ubiquitous}. Therefore, we have $\pmb{x}=\pmb{\Upsilon}\pmb{s}$, where $\pmb{s}=\text{vec}(\pmb{S})\in\mathbb{C}^{M_dN_t\times 1}$. Consequently, the end-to-end input-output relationship can be rewritten as 
\begin{align}\label{Eq10}
	\pmb{y}=\pmb{C}\pmb{s}+\pmb{n},
\end{align}
where $\pmb{C}=\pmb{H}\pmb{\Upsilon}$ represents the equivalent channel matrix, and the equivalent input symbol vector can be expressed as $\pmb{s}=[\pmb{s}_0^T,\pmb{s}_1^T,\ldots,\pmb{s}_{M_d-1}^T]^T$. The probability density function (PDF) of $p(\pmb{y}|\pmb{s})$ is
\begin{align}\label{Eq10-1}
	p(\pmb{y}|\pmb{s})=\frac{1}{(\pi \sigma^2)^{M_dN_r}}\exp\left(-\frac{||\pmb{y}-\pmb{C}\pmb{s}||^2}{\sigma^2}\right).
\end{align}

Note that the sub-vector $\pmb{s}_{m_d}$ of $\pmb{s}$ is given by the $m_d$th column of $\pmb{S}$ in \eqref{Eq1}, which hence only has a single non-zero element. Therefore, there are overall $C=2^{M_dL_1}=N_t^{M_d}$ TAPs $\mathcal{Q}=\{\mathcal{Q}_1,\ldots,\mathcal{Q}_C\}$, and the $c$th TAP is denoted as $\mathcal{Q}_c=\{\mathcal{Q}_{c,0},\ldots,\mathcal{Q}_{c,M_d-1}\}$, where we have $\mathcal{Q}_{c,m_d}\in\mathbb{Z}_{+}^{N_tM_d}$ for $m_d=0,\ldots,M_d-1$ and $\mathbb{Z}_{+}^{N_tM_d}$ denotes the real integer set $\{1,\ldots,N_tM_d\}$. Given a TAP, the indices of the activated TAs are denoted as $\mathcal{I}=\mathcal{Q}_c\subset\mathcal{Q}$. Furthermore, the APM symbol vector $\pmb{s}_{\text{D}}=[s_{\text{D}}(0),\ldots,s_{\text{D}}(M_d-1)]^T\in\mathcal{A}^{M_d\times 1}$ has $Q^{M_d}$ realizations.
\section{Detection Algorithms and DCMC Capacity}\label{Section3}
In this section, we commence by detailing the optimum MLD of our SM-OTFS system. Explicitly, the complexity of MLD may be excessive when the value of $L_b$ is high. Therefore, we propose a low-complexity near-ML DOSCD. Moreover, the complexity analysis of the detectors is provided. Finally, the DCMC capacity of the SM-OTFS system is derived.
\subsection{Maximum Likelihood Detector}\label{Section3-1}
According to the analysis in Section \ref{Section2}, the total number of the realisations of $\pmb{s}$ can be expressed as $|\Omega|=2^{L_b}=(N_tQ)^{M_d}$, where $\Omega$ is the set of candidates of $\pmb{s}$. Under the condition that all the candidates are independent and equiprobable, the optimal MLD can be formulated as
\begin{align}\label{Eq10-1}
	\pmb{s}^{\text{ML}}=\argmin_{\pmb{s}\in\Omega}\left\{||\pmb{y}-\pmb{C}\pmb{s}||^2\right\}.
\end{align}
\subsection{Distance-based Ordering Subspace Check Detector}\label{Section3-2}
One of the objectives of DOSCD is to detect the APM symbols and TAPs in different subspaces separately. Given the TAP $\mathcal{I}$, we have $\pmb{s}=\pmb{\Upsilon}_{\mathcal{I}}\pmb{s}_{\text{D}}$, where $\pmb{\Upsilon}_{\mathcal{I}}$ is the $(M_dN_t\times M_d)$-dimensional element mapping matrix based on $\mathcal{I}$. Then the input-output relationship of \eqref{Eq10} can be rewritten as
\begin{align}\label{Eq11}
	\pmb{y}&=\pmb{C}\pmb{s}+\pmb{n}\nonumber\\
	&=\pmb{C}\pmb{\Upsilon}_{\mathcal{I}}\pmb{s}_{\text{D}}+\pmb{n}\nonumber\\
	&=\pmb{C}_{\mathcal{I}}\pmb{s}_{\text{D}}+\pmb{n}.
\end{align}

Let us first employ the linear minimum mean square error (LMMSE) detector to obtain the soft estimate of $\pmb{s}$, yielding 
\begin{align}\label{Eq12}
	\tilde{\pmb{s}}=\left(\pmb{C}^H\pmb{C}+\frac{1}{\gamma_s}\pmb{I}_{N_tM_d}\right)^{-1}\pmb{C}^H\pmb{y},
\end{align}
where the average signal-to-noise ratio (SNR) per symbol is given by $\gamma_s=1/(N_t\sigma^2)$. Then the hard decision based on $\tilde{\pmb{s}}$ can be obtained by leveraging the simple element-wise rounding-based demodulation \cite{6679367}, which is given by as $\hat{\pmb{s}}=\mathbb{Q}(\tilde{\pmb{s}})$. Hence, the distance between $\tilde{\pmb{s}}$ and $\hat{\pmb{s}}$ can be expressed as $\pmb{d}=[d(0),\ldots,d(N_tM_d-1)]^T$, where we have $d(i)=|\tilde{s}(i)-\hat{s}(i)|^2$ for $i=0,\ldots,N_tM_d-1$. After obtaining the distances $d(i)$ of all the possible indices, the reliabilities of the elements in $\tilde{\pmb{s}}$ can be measured. Explicitly, the soft estimates $\tilde{s}(i)$ corresponding to the smaller values of $d(i)$ are more reliable. Here, we emphasize that there are only $M_d$ non-zero elements in the equivalent transmitted symbol vector $\pmb{s}$, whose indices depend on the correct TAP.

Specifically, for the $c$th TAP $\mathcal{Q}_c=\{\mathcal{Q}_{c,0},\ldots,\mathcal{Q}_{c,M_d-1}\}$ and the distance elements $d(i)$ for $i=0,\ldots,N_tM_d-1$, the sum of the corresponding distance elements $d(\mathcal{Q}_{c,m_d}-1)$ for $m_d=0,\ldots,M_d-1$ is calculated, yielding the reliability metric for the $c$th TAP as
\begin{align}\label{Eq13}
	\lambda_c=\sum_{m_d=0}^{M_d-1}d(\mathcal{Q}_{c,m_d}-1),\quad c=1,\ldots,C.
\end{align}
Now, let us sort all the reliability metrics in ascending order to form an ordering set as
\begin{align}\label{Eq14}
		\{{i_1},\ldots,{i_C}\}\quad\text{subject to }{\lambda}_{i_1}\leqslant\ldots\leqslant{\lambda}_{i_C},
\end{align}
where $i_c\in\{1,\ldots,C\}$ and $i_j\neq i_q$, $\forall j\neq q$. Based on the above analysis, it is plausible that the TAP yielding a smaller value of $\lambda_{i_c}$ has a higher probability of being the correct TAP, which becomes more evident, as SNR increases \cite{7583706}.

To improve the detection performance, let us consider testing the first $T_d$ TAPs in the spirit of \eqref{Eq10-1} according to the ordering reliability metrics in \eqref{Eq14}. For this purpose, first, based on \eqref{Eq11}, the least square estimation is executed to obtain the soft estimates of the APM symbols, which can be expressed as $\breve{\pmb{s}}_\text{D}^t=\pmb{C}_{\mathcal{Q}^t}^\dagger\pmb{y}$ for $t=1,\ldots,T_d$, where $\pmb{C}_{\mathcal{Q}^t}^\dagger$ denotes the Moore-Penrose pseudoinverse matrix of $\pmb{C}_{\mathcal{Q}^t}$ \cite{7583706}. Then, the APM symbols delivered by the $t$th TAP can be detected by invoking the simple element-wise rounding-based demodulation approach of \cite{6679367}, yielding 
	$\pmb{s}_{\text{D}}^t=\mathbb{Q}(\breve{\pmb{s}}_\text{D}^t)$. After this detection, the residual error can be expressed as
	\begin{align}\label{Eq16}
		\epsilon^t=|\pmb{y}-\pmb{C}_{\mathcal{Q}^t}\pmb{s}_{\text{D}}^t|,\quad t=1,\ldots,T_d.
	\end{align}
Finally, the index of the optimal TAP can be formulated as 
\begin{align}
	\hat{t}=\underset{t\in\{1,\ldots,T_d\}}\argmin \epsilon^t.
\end{align}
Consequently, the final detected TAP and the APM symbols conveyed are given by $\mathcal{I}^{\text{DOSCD}}=\mathcal{Q}^{\hat{t}},\quad \pmb{s}_{\text{D}}^{\text{DOSCD}}=\pmb{s}_{\text{D}}^{\hat{t}}$.

In summary, our DOSCD is described in Algorithm \ref{alg1}.
\begin{algorithm}[h!]
\caption{{Distance-based Ordering Subspace Check Detector}}
\label{alg1}
\begin{algorithmic}[1]
    \Require
      $\pmb{y}$, $\pmb{C}$, $\gamma_s$ and $\mathcal{Q}$.
      \State \textbf{Preparation}: Set the maximum number $T_d$ of candidates to be tested.
    \State Employ the LMMSE detection based on \eqref{Eq12} as
    \State $\tilde{\pmb{s}}=\left(\pmb{C}^H\pmb{C}+\frac{1}{\gamma_s}\pmb{I}_{N_tM_d}\right)^{-1}\pmb{C}^H\pmb{y}$.
    \State Obtain the hard decision of $\tilde{\pmb{s}}$ as $\hat{\pmb{s}}=\mathbb{Q}(\tilde{\pmb{s}})$.
    \State Based on $\tilde{\pmb{s}}$ and $\hat{\pmb{s}}$, compute the distance vector $\pmb{d}=[d(0),\ldots,d(N_tM_d-1)]^T$,
    \Statex\ where $d(i)=|\tilde{s}(i)-\hat{s}(i)|^2$ for $i=0,\ldots,N_tM_d-1$.
    \For{$c=1$ to $C$}
    \State Calculate the reliability metric based on the $c$th TAP $\mathcal{Q}_c=\{\mathcal{Q}_{c,0},\ldots,\mathcal{Q}_{c,M_d-1}\}$ as 
    \State $\lambda_c=\sum_{m_d=0}^{M_d-1}d(\mathcal{Q}_{c,m_d}-1)$.
    \EndFor
    \State Sort the reliability metrics in ascending order as
    \State $\{{i_1},\ldots,{i_C}\}\quad\text{subject to }{\lambda}_{i_1}\leqslant\ldots\leqslant{\lambda}_{i_C}$.
    \For{$t=1$ to $T_d$}
    \State Collect the TA activation scheme $\mathcal{Q}^t$ according to $\lambda_{i_t}$.
    \State Carry out the least square estimation as $\breve{\pmb{s}}_\text{D}^t=\pmb{C}_{\mathcal{Q}^t}^\dagger\pmb{y}$.
    \State Calculate the residual error as $\epsilon^t=|\pmb{y}-\pmb{C}_{\mathcal{Q}^t}\pmb{s}_{\text{D}}^t|$.
    \EndFor
    \State Find the index of the optimal TAP as
    \State $\hat{t}=\underset{t\in\{1,\ldots,T_d\}}\argmin \epsilon^t$. 
\State \textbf{Output} $\mathcal{I}^{\text{DOSCD}}=\mathcal{Q}^{\hat{t}}$ and $\pmb{s}_{\text{D}}^{\text{DOSCD}}=\pmb{s}_{\text{D}}^{\hat{t}}$.
\end{algorithmic}
\end{algorithm}
\vspace{-1em}
\subsection{Complexity Analysis}\label{Section3-4}
Firstly, according to the analysis in Section \ref{Section3-1}, the MLD searches all the candidates in $\Omega$, at the complexity order of $\mathcal{O}[(N_tQ)^{M_d}]$.

Then, the resultant complexity of MPD is on the order of $\mathcal{O}[M^3N^3(N_r^2N_t+N_t^2N_r)N_tQT_\text{MP}/N_t^2]$, where $T_\text{MP}$ denotes the number of iterations in MPD \cite{10039709}.

Finally, based on our analysis in Section \ref{Section3-2}, the complexity of a single DOSCD iteration is on the order of $\mathcal{O}({M_d})$. Therefore, we can infer that the complexity order of the overall DOSCD is given by $\mathcal{O}(T_d{M_d})$, since there are $T_d$ iterations. The worst-case scenario is, when then all the $T_{\text{max}}=N_t^{M_d}$ TAPs are tested. However, as the simulation results of Section \ref{Section4} will show, the DOSCD is capable of attaining a near-ML BER performance for $T_d\ll N_t^{M_d}$.
\subsection{DCMC Capacity}\label{Section3-5}
Based on Section \ref{Section3-1}, we denote all the realisations of $\pmb{s}$ as $\Omega=\{\bar{\pmb{s}}_1,\ldots,\bar{\pmb{s}}_{2^{L_b}}\}$. The DCMC capacity of the SM-OTFS system can be formulated as \cite{1608632}
\begin{align}\label{Eq22}
	C_\text{D}&=\frac{1}{M_d}\underset{p(\bar{\pmb{s}}_1)\ldots p(\bar{\pmb{s}}_{2^{L_b}})}\max\sum_{i=1}^{2^{L_b}}\int_{-\infty}^{\infty}\ldots\int_{-\infty}^{\infty}p(\pmb{y}|\bar{\pmb{s}}_i)p(\bar{\pmb{s}}_i)\nonumber\\
	&\times\log_2\left[\frac{p(\pmb{y}|\bar{\pmb{s}}_i)}{\sum_{j=1}^{2^{L_b}}p(\pmb{y}|\bar{\pmb{s}}_j)p(\bar{\pmb{s}}_j)}\right]d\pmb{y},
\end{align}
where $p(\pmb{y}|\bar{\pmb{s}}_i)$ is given by \eqref{Eq10-1}, when $\bar{\pmb{s}}_i$ is transmitted. It can be readily shown that \eqref{Eq22} attains its maximum value in the case of $p(\pmb{y}|\bar{\pmb{s}}_i)=1/2^{L_b}$, $\forall i$. Therefore, the DCMC capacity of the SM-OTFS system can be shown to be
\begin{align}\label{Eq23}
C_\text{D}=\frac{1}{M_d}\left\{L_b-\frac{1}{2^{L_b}}\sum_{i=1}^{2^{L_b}}\mathbb{E}\left[\log_2\sum_{j=1}^{2^{L_b}}\exp\left(\Psi\right)\right]\right\},
	\end{align}
where we have $\Psi={\left[-||\pmb{C}(\bar{\pmb{s}}_i-\bar{\pmb{s}}_j)+\pmb{n}||^2+||\pmb{n}||^2\right]}/{\sigma^2}$ by substituting \eqref{Eq10-1} into \eqref{Eq23}. It can be readily shown that the DCMC capacity is upper bounded by 
\begin{align}\label{Eq24}
	C_\text{D,max}=\frac{L_b}{M_d}=\log_2(N_tQ).
	\end{align}
We note that the closed-form DCMC capacity expression is computationally intractable, since there are multidimensional integrals and summations of $2^{L_b}$ exponential functions in \eqref{Eq22} and \eqref{Eq23}, respectively. Moreover, it can be seen from \eqref{Eq23} that the DCMC capacity is dependent on all the realisations of $\bar{\pmb{s}}_i$. Therefore, in general, the classic Monte Carlo averaging method is exploited to compute the DCMC capacity $C_\text{D}$ \cite{1608632}.
\section{Performance Analysis}\label{Section4}
In this section, we provide simulation results for characterizing the performance of the SM-OTFS system and the proposed detectors. The carrier frequency and subcarrier spacing are set to $f_c=4$ GHz and $\Delta f=15$ kHz, respectively. For the sake of comparison, the conventional single-input-multiple-output (SIMO)-OTFS specified as $(N_r,Q)$ is used as our benchmark, yielding a rate of $R=\log_2 Q$ bits/s/Hz. We assume that the SM-OTFS and SIMO-OTFS systems have $M=8$ subcarriers and $N=4$ TSs. A four-path (i.e., $P=4$) doubly-selective channel is considered, whose maximum normalized Doppler and delay shifts are $k_\text{max}=N-1$ and $l_\text{max}=M-1$ \cite{8686339}, respectively. The normalized shifts of the $i$th path are given as $k_i\in U[-k_\text{max},k_\text{max}]$ and $l_i\in U[0,l_\text{max}]$, respectively, where $U[a,b]$ denotes the uniform distribution in the interval $[a,b]$.

\begin{figure}[htbp]
\centering
\vspace{-0.5em}
\centering
\includegraphics[width=\linewidth]{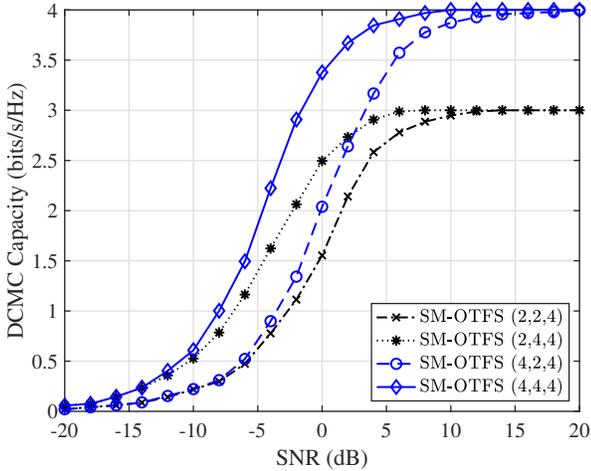}
\caption{The DCMC capacity of the SM-OTFS $(N_t,N_r,4)$ systems parameterized by $N_t$ and $N_r$.}
\label{Figure2}
\end{figure}

In Fig. \ref{Figure2}, the DCMC capacity of SM-OTFS $(N_t,N_r,4)$ systems having different settings of $N_t$ and $N_r$ are depicted. It can be observed from Fig. \ref{Figure2} that the SM-OTFS systems having $N_t=4$ are capable of attaining a higher DCMC capacity than their $N_t=2$ counterparts. This is because the signal dimensionality becomes higher as the value of $N_t$ escalates in our SM-OTFS system. Furthermore, given the number of TAs, the DCMC capacity converges to the upper-bound $C_\text{D,max}$ in \eqref{Eq24} that is independent of the number of RAs, implying that although a higher receiver diversity order can be obtained and a higher capacity can be achieved in the low-SNR region by employing more RAs, the DCMC capacity will not be increased in the high-SNR region. The above-mentioned observations are consistent with the conclusions of \cite{1608632}.

\begin{figure}[htbp]
\vspace{-1em}
\centering
\includegraphics[width=\linewidth]{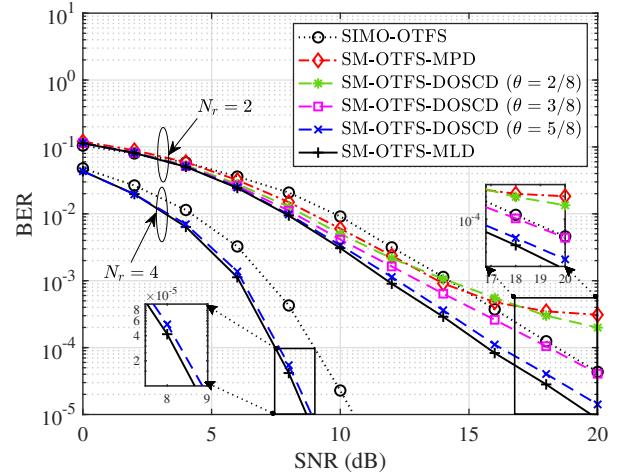}
\caption{BER performance of the SIMO-OTFS $(N_r,8)$ using MLD, and the SM-OTFS $(2,N_r,4)$ using DOSCD parameterized by DOSCD iterations, as well as MPD \cite{10039709} and MLD when operating at 3 bits/s/Hz.}
\label{Figure3}
\end{figure}

Fig. \ref{Figure3} investigates the BER performance of the SIMO-OTFS system $(N_r,8)$ using MLD, and of SM-OTFS $(2,N_r,4)$ using DOSCD at different numbers of iterations, as well as MPD \cite{10039709} and MLD at $R=3$ bits/s/Hz, when $N_r=2$ or $4$ RAs are employed. The number of MPD iterations is $T_\text{MP} = 10$. Explicitly, $T_d=\theta T_{\text{max}}$ DOSCD iterations are employed along with $\theta=2/8$, $3/8$ and $5/8$, respectively. From Fig. \ref{Figure3}, we have the following observations. Firstly, the BER performance of SM-OTFS is better than that of the SIMO-OTFS system for both the $N_r=2$ and $N_r=4$ cases. Specifically, given a BER of $10^{-4}$, our SM-OTFS using MLD is capable of attaining about $3$ dB and $1.5$ dB gains compared to the conventional SIMO-OTFS scheme for $N_r=2$ and $N_r=4$, respectively. This is because SM-OTFS can achieve extra spatial diversity gain, while employing a lower-order modulation scheme compared to the SIMO-OTFS operating at a rate of $R$. Secondly, the higher the value of $T_d$, the better the BER performance attained by the DOSCD. This can be explained by the fact that when more TAPs are tested in the DOSCD, better detection performance can be achieved. More specifically, for a BER of $10^{-4}$, the DOSCD employing $\theta=5/8$ attains about 2 dB gain over its $\theta=3/8$ counterpart in the case of $N_t=2$. Moreover, given $N_t=2$, the SM-OTFS associated with DOSCD using both $\theta=3/8$ and $\theta=5/8$ is capable of attaining better BER performance than the SIMO-OTFS using MLD. Furthermore, given $N_r=2$ and a BER of $10^{-4}$, the DOSCD associated with $\theta=5/8$ attains a near-ML BER with a performance gap of 0.5 dB. In addition, given a BER of $10^{-3}$, the proposed DOSCD using $\theta=5/8$ is capable of attaining about $2.5$ dB gain compared to the MPD with $N_r=2$. Explicitly, it can be observed that the MPD's BER curve exhibits an error floor at a BER of $3\times 10^{-4}$, which is consistent with \cite{10039709}. This is because the Gaussian interference assumption based on the central-limit theorem is inaccurate \cite{9508141}. Finally, the BER performance of the DOSCD with $\theta=5/8$ is nearly identical to that of the MLD in the case of $N_r=4$, which is a trend reminiscent of massive MIMO uplink detection, as explained by the channel hardening phenomenon \cite{9729571}.
\begin{figure}[htbp]
\centering
\vspace{-0.3cm}
\centering
\includegraphics[width=\linewidth]{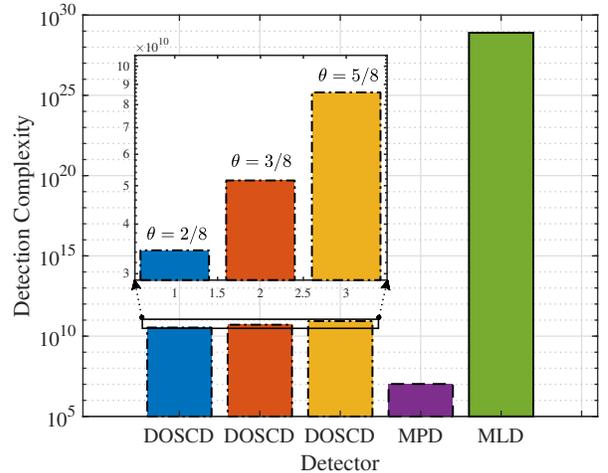}
\caption{The detection complexity of SM-OTFS $(2,2,4)$ using DOSCD with different values of $\theta$, MPD \cite{10039709} and MLD operating at 3 bits/s/Hz.}
\vspace{-0.7em}
\label{Figure4}
\end{figure}

The detection complexity of the SM-OTFS $(2,2,4)$ system employing the DOSCD along with different values of $\theta$, MPD \cite{10039709} and MLD are compared in Fig. \ref{Figure4}. We observe that the DOSCD using $\theta=5/8$ achieves a significant (i.e., 28 orders of magnitude) complexity reduction over the MLD. This can be explained by the fact that the DOSCD invokes simple symbol-wise detection, while the reliability evaluation step of \eqref{Eq14} further reduces the number of DOSCD iterations. Moreover, the DOSCD using $\theta=2/8$ and $\theta=3/8$ attains about 60\% and 40\% complexity reduction over the $\theta=5/8$ case, respectively, since more TAPs are tested along with higher values of $\theta$. Furthermore, although the DOSCD using $\theta=5/8$ is capable of attaining the best BER performance compared to other DOSCD cases, this improvement is attained at the cost of higher complexity. In addition, the complexity of our DOSCD is higher than that of the MPD, which is the cost of attaining better BER performance. Finally, it can be observed based on Fig. \ref{Figure3} and Fig. \ref{Figure4} that the DOSCD provides satisfactory near-ML BER performance at reduced complexity.
\section{Summary and Conclusions}\label{Section5}
The input-output relationship of the SM-OTFS system has been derived for transmission over doubly-selective channels. By exploiting the properties of SM, a symbol mapper has been introduced for deriving the input-output relationship of SM-OTFS. Furthermore, a low-complexity detector has been proposed for SM-OTFS systems. Explicitly, the reliabilities of TAPs have been evaluated and the TAPs as well as APM symbols have been detected separately in the DOSCD. Our simulation results have shown that the proposed SM-OTFS system provides a better BER performance than the conventional SIMO-OTFS, and the proposed DOSCD is capable of striking a compelling BER vs. complexity trade-off.
\renewcommand{\refname}{References}
\mbox{} 
\nocite{*}
\bibliographystyle{IEEEtran}
\bibliography{TVTL_Zeping.bib}

\begin{thebibliography}{10}
\providecommand{\url}[1]{#1}
\csname url@samestyle\endcsname
\providecommand{\newblock}{\relax}
\providecommand{\bibinfo}[2]{#2}
\providecommand{\BIBentrySTDinterwordspacing}{\spaceskip=0pt\relax}
\providecommand{\BIBentryALTinterwordstretchfactor}{4}
\providecommand{\BIBentryALTinterwordspacing}{\spaceskip=\fontdimen2\font plus
\BIBentryALTinterwordstretchfactor\fontdimen3\font minus
  \fontdimen4\font\relax}
\providecommand{\BIBforeignlanguage}[2]{{%
\expandafter\ifx\csname l@#1\endcsname\relax
\typeout{** WARNING: IEEEtran.bst: No hyphenation pattern has been}%
\typeout{** loaded for the language `#1'. Using the pattern for}%
\typeout{** the default language instead.}%
\else
\language=\csname l@#1\endcsname
\fi
#2}}
\providecommand{\BIBdecl}{\relax}
\BIBdecl

\bibitem{6678765}
M.~Di~Renzo, H.~Haas, A.~Ghrayeb, S.~Sugiura, and L.~Hanzo, ``Spatial
  modulation for generalized {MIMO}: Challenges, opportunities, and
  implementation,'' \emph{Proceedings of the IEEE}, vol. 102, no.~1, pp.
  56--103, 2014.

\bibitem{4382913}
R.~Y. Mesleh, H.~Haas, S.~Sinanovic, C.~W. Ahn, and S.~Yun, ``Spatial
  modulation,'' \emph{IEEE Transactions on Vehicular Technology}, vol.~57,
  no.~4, pp. 2228--2241, 2008.

\bibitem{9729571}
J.~An, C.~Xu, Y.~Liu, L.~Gan, and L.~Hanzo, ``The achievable rate analysis of
  generalized quadrature spatial modulation and a pair of low-complexity
  detectors,'' \emph{IEEE Transactions on Vehicular Technology}, vol.~71,
  no.~5, pp. 5203--5215, 2022.

\bibitem{6679367}
R.~Rajashekar, K.~Hari, and L.~Hanzo, ``Reduced-complexity {ML} detection and
  capacity-optimized training for spatial modulation systems,'' \emph{IEEE
  Transactions on Communications}, vol.~62, no.~1, pp. 112--125, 2014.

\bibitem{6376045}
Q.~Tang, Y.~Xiao, P.~Yang, Q.~Yu, and S.~Li, ``A new low-complexity near-{ML}
  detection algorithm for spatial modulation,'' \emph{IEEE Wireless
  Communications Letters}, vol.~2, no.~1, pp. 90--93, 2013.

\bibitem{6784532}
Y.~Xiao, Z.~Yang, L.~Dan, P.~Yang, L.~Yin, and W.~Xiang, ``Low-complexity
  signal detection for generalized spatial modulation,'' \emph{IEEE
  Communications Letters}, vol.~18, no.~3, pp. 403--406, 2014.

\bibitem{9468877}
M.~Saad, H.~Hijazi, A.~C.~A. Ghouwayel, F.~Bader, and J.~Palicot, ``Low
  complexity quasi-optimal detector for generalized spatial modulation,''
  \emph{IEEE Communications Letters}, vol.~25, no.~9, pp. 3003--3007, 2021.

\bibitem{7925924}
R.~Hadani, S.~Rakib, M.~Tsatsanis, A.~Monk, A.~J. Goldsmith, A.~F. Molisch, and
  R.~Calderbank, ``Orthogonal time frequency space modulation,'' in \emph{2017
  IEEE Wireless Communications and Networking Conference (WCNC)}, 2017, pp.
  1--6.

\bibitem{9508932}
Z.~Wei, W.~Yuan, S.~Li, J.~Yuan, G.~Bharatula, R.~Hadani, and L.~Hanzo,
  ``Orthogonal time-frequency space modulation: A promising next-generation
  waveform,'' \emph{IEEE Wireless Communications}, vol.~28, no.~4, pp.
  136--144, 2021.

\bibitem{9508141}
L.~Xiang, Y.~Liu, L.-L. Yang, and L.~Hanzo, ``Gaussian approximate message
  passing detection of orthogonal time frequency space modulation,'' \emph{IEEE
  Transactions on Vehicular Technology}, vol.~70, no.~10, pp. 10\,999--11\,004,
  2021.

\bibitem{8686339}
G.~D. Surabhi, R.~M. Augustine, and A.~Chockalingam, ``On the diversity of
  uncoded {OTFS} modulation in doubly-dispersive channels,'' \emph{IEEE
  Transactions on Wireless Communications}, vol.~18, no.~6, pp. 3049--3063,
  2019.

\bibitem{yuan2021iterative}
Z.~Yuan, F.~Liu, W.~Yuan, Q.~Guo, Z.~Wang, and J.~Yuan, ``Iterative detection
  for orthogonal time frequency space modulation with unitary approximate
  message passing,'' \emph{IEEE transactions on wireless communications},
  vol.~21, no.~2, pp. 714--725, 2021.

\bibitem{9521176}
Y.~Yang, Z.~Bai, K.~Pang, P.~Ma, H.~Zhang, X.~Yang, and D.~Yuan, ``Design and
  analysis of spatial modulation based orthogonal time frequency space
  system,'' \emph{China Communications}, vol.~18, no.~8, pp. 209--223, 2021.

\bibitem{9718240}
T.~Wang, S.~Fan, H.~Chen, Y.~Xiao, X.~Guan, and W.~Song, ``Generalized
  approximate message passing detector for {GSM-OTFS} systems,'' \emph{IEEE
  Access}, vol.~10, pp. 22\,997--23\,007, 2022.

\bibitem{8424569}
P.~Raviteja, K.~T. Phan, Y.~Hong, and E.~Viterbo, ``Interference cancellation
  and iterative detection for orthogonal time frequency space modulation,''
  \emph{IEEE Transactions on Wireless Communications}, vol.~17, no.~10, pp.
  6501--6515, 2018.

\bibitem{10039709}
S.~Li, L.~Xiao, X.~Zhang, L.~Li, and T.~Jiang, ``Spatial multiplexing aided
  {OTFS} with index modulation,'' \emph{IEEE Transactions on Vehicular
  Technology}, pp. 1--6, 2023.

\bibitem{9590508}
S.~Srivastava, R.~K. Singh, A.~K. Jagannatham, and L.~Hanzo, ``Bayesian
  learning aided simultaneous row and group sparse channel estimation in
  orthogonal time frequency space modulated {MIMO} systems,'' \emph{IEEE
  Transactions on Communications}, vol.~70, no.~1, pp. 635--648, 2022.

\bibitem{9906092}
M.~Mohammadi, H.~Q. Ngo, and M.~Matthaiou, ``Cell-free massive {MIMO} meets
  {OTFS} modulation,'' \emph{IEEE Transactions on Communications}, vol.~70,
  no.~11, pp. 7728--7747, 2022.

\bibitem{bjornson2015massive}
E.~Bj{\"o}rnson, E.~G. Larsson, and M.~Debbah, ``Massive {MIMO} for maximal
  spectral efficiency: How many users and pilots should be allocated?''
  \emph{IEEE Transactions on Wireless Communications}, vol.~15, no.~2, pp.
  1293--1308, 2015.

\bibitem{9891774}
S.~Srivastava, R.~K. Singh, A.~K. Jagannatham, and L.~Hanzo, ``Delay-doppler
  and angular domain {4D}-sparse {CSI} estimation in {OTFS} aided {MIMO}
  systems,'' \emph{IEEE Transactions on Vehicular Technology}, vol.~71, no.~12,
  pp. 13\,447--13\,452, 2022.

\bibitem{10042436}
S.~Li, J.~Yuan, P.~Fitzpatrick, T.~Sakurai, and G.~Caire, ``Delay-doppler
  domain tomlinson-harashima precoding for {OTFS}-based downlink {MU-MIMO}
  transmissions: Linear complexity implementation and scaling law analysis,''
  \emph{IEEE Transactions on Communications}, pp. 1--1, 2023.

\bibitem{van2000ubiquitous}
C.~F. Van~Loan, ``The ubiquitous {Kronecker} product,'' \emph{Journal of
  computational and applied mathematics}, vol. 123, no. 1-2, pp. 85--100, 2000.

\bibitem{7583706}
H.~Zhang, L.-L. Yang, and L.~Hanzo, ``Compressed sensing improves the
  performance of subcarrier index-modulation-assisted {OFDM},'' \emph{IEEE
  Access}, vol.~4, pp. 7859--7873, 2016.

\bibitem{1608632}
S.~X. Ng and L.~Hanzo, ``On the {MIMO} channel capacity of multidimensional
  signal sets,'' \emph{IEEE Transactions on Vehicular Technology}, vol.~55,
  no.~2, pp. 528--536, 2006.

\end{thebibliography}

\end{document}